\documentclass{ptapap}
\usepackage{fixltx2e}

\author{Melissa Munoz}[QU]
\author{Gregg Wade}[RMC]
\author{Ya\"{e}l Naz\'{e}}[Liege]
\author{Stefano Bagnulo}[IAU]
\author{Joachim Puls}[ISM]
\affil[QU]{Queen's University\\
  99 University Ave, Kingston, Ontario, Canada, 16msm5@queensu.ca}
\affil[RMC]{Royal Military College of Canada\\
	13 General Crerar Crescent, Kingston, Ontario, Canada}
\affil[Liege]{Universit\'{e} de Li\`{e}ge\\
	Place du 20 Août 7, 4000 Liège, Belgium}
\affil[IAU]{Armagh Observatory\\
College Hill, Armagh BT61 9DG, Northern Ireland, United Kingdom}
\affil[ISM]{Universit\"{a}tssternwarte \\
	 Scheinerstr. 1, D-81679 M\"{u}nchen, Germany}
\title{The Of?p stars of the Magellanic Clouds: Are 
	they strongly magnetic?}

\begin{document}

\maketitle

\begin{abstract}
All known Galactic Of?p stars have been shown to host strong, organized, magnetic fields. Recently, five Of?p stars have been discovered in the Magellanic Clouds. They posses photometric \citep{Naze} and spectroscopic \citep{Walborn} variability compatible with the Oblique Rotator Model (ORM). However, their magnetic fields have yet to be directly detected. We have developed an algorithm allowing for the synthesis of photometric observables based on the Analytic Dynamical Magnetosphere (ADM) model of \citet{Owocki}. We apply our model to OGLE photometry in order to constrain their magnetic geometries and and surface dipole strengths. We predict that the field strengths for some of these candidate extra-Galactic magnetic stars may be within the detection limits of the FORS2 instrument.
\end{abstract}

\section{Introduction}
Of?p stars correspond to a small subset of O-type stars that host unusual spectral properties. In particular, their spectra show  N III $\lambda\lambda$ 4634-41 lines in emission with strengths comparable to their neighboring C III $\lambda$ 4686 emission lines. Decades after their introduction by Walborn (1972), they were further shown to display phase-locked variability  \citep[see][]{Naze2008}. Thanks to the collaborative efforts of the MiMeS (Magnetism in Massive Stars) survey, it has recently been shown that all known Galactic Of?p stars are in fact magnetic \citep[e.g.][]{Grunhut}. The periodic spectroscopic and photometric variability is consistent with the Oblique Rotator Model \citep[ORM,][]{ORM}, where the observed variabilities are in fact rotationally induced modulations.

Recently, the first candidate extra-Galactic Of?p stars have been discovered in the Small and Large Magellanic Clouds \citep{Walborn}. They also possess coherent phase dependent spectroscopic and photometric variability similar to the Galactic sample of Of?p stars. It is thus highly suspected that the Magellanic Of?p stars host magnetic fields as well. \citet{Bagnulo} made a first attempt to detect these proposed magnetic fields using the FORS2 instrument at VLT. No magnetic fields were detected, but due to poor weather the precision obtained was much worse than projected.

Therefore, we propose an indirect method to infer the magnetic field properties of magnetic massive stars via modeling of the photometric variations caused by the rotation-phase-dependent scattering of their photospheric light by their magnetospheres. Section \ref{sec:model} will describe the numerical method allowing for the light curve synthesis. In Section  \ref{sec:obs}, we will first apply our algorithm to archival Hipparcos photometry of the thoroughly-studied   Galactic Of?p star, HD 191612 \citep{Wade2011}  as a proof of concept, and then to the recent OGLE photometry of the Magellanic Of?p stars \citep{Naze}. Finally, we will summarize our results in the last section. 


\section{Model light curves}\label{sec:model}
Magnetic massive stars host stable, large scale magnetic fields that confine their winds symmetrically about the equator. If, for such stars, the magnetic equator is tilted with respect to the stellar rotation axis, they will manifest phase-locked observable quantities, namely, the photometric brightness due to the rotation of an asymmetric wind structure. Section \ref{subsec:algo} will describe the numerical prescription allowing for the light curve synthesis, while Section \ref{subsec:params} will demonstrate some light curve models as part of a parameter space study. 

\subsection{The algorithm}\label{subsec:algo}
For a wind dominated by electron scattering (which is appropriate for hot massive stars), the photometric variability will be modulated by the amount of plasma occulting the star's surface along the line-of-sight of an observer. This is parametrized via the $\alpha$ angle corresponding to the inclination of the magnetic equator with respect to the observer's line-of-sight. The projection of such an angle is 
\begin{equation} \label{eq:alpha}
\cos \alpha = \cos \phi \sin \beta + \cos \beta \cos i,
\end{equation}
where $i$ is the inclination angle (angle between the observer's line-of-sight and the rotation axis), $\beta$ is the magnetic obliquity (angle between the magnetic axis and the rotation axis) and $\phi$ is the rotational phase. 

At each rotational phase, the differential magnitude due to an optically thin single electron scattering wind is given by 
\begin{equation}\label{eq:mag}
	\Delta m = \Delta m_0 + (2.5 \log e) \tau,
\end{equation}
where $\Delta m_0$ is a constant magnitude shift and $\tau$ is the Thomson scattering optical depth. For an arbitrarily placed observer, the line-of-sight optical depth will be related to the magnetosphere density, $\rho$, as follows
\begin{equation}\label{eq:tau}
	\tau = \frac{\alpha_e \sigma_e}{m_p} \int \rho dz,
\end{equation}
where $\sigma_e$ is the Thomson cross section, $\alpha_e$ is the free electron baryon mass and $m_p$ is the mass of a proton. For a completely ionized gas composed of helium, that is suitable for hot massive stars, $\alpha_e=0.5$. 

The density structure surrounding the star is provided by the Analytic Dynamical Magnetosphere (ADM) model  from \citet{Owocki}. The ADM model is capable of characterizing the density, temperature and wind flow structures via analytical prescriptions. For simplicity, it assumes a dipolar field geometry which is observationally consistent with the well-known Galactic Of?p stars \citep{Grunhut}. Traditionally, magnetosphere models were obtained using computationally expensive MHD simulations. As the ADM model is vastly more computationally efficient and was shown to be in good agreement with more sophisticated MHD simulations \citep{Owocki}, we have chosen to implement the ADM formalism.

\subsection{A parameter space study} \label{subsec:params}
The simplicity of the light curve model allows us to readily conduct a parameter space study. We first need to consider the entire set of free parameters. The ADM model alone requires six input parameters: the mass-loss rate, wind terminal velocity, stellar mass, radius, effective temperature and magnetic dipole field strength. The first five can generally be well restrained by spectral classification or spectroscopic modeling. We will therefore consider a synthetic star with fixed stellar and wind properties similar to HD191612, an Of?p prototype. In addition to the dipole field strength ($B_d$), two free parameters are needed in order to specify the geometry of the magnetosphere: the inclination angle ($i$) and magnetic obliquity ($\beta$). 

Note that eq. \ref{eq:alpha} is symmetric under $i$ and $\beta$. In other words, these two angles are degenerate and they can be interchanged without producing a change in light curve shape. We will thus only consider the sum of these two angles in the parameter space study. 
\begin{figure}
	\includegraphics[width=\textwidth, trim={3cm 1.5cm 3cm 1.5cm}]{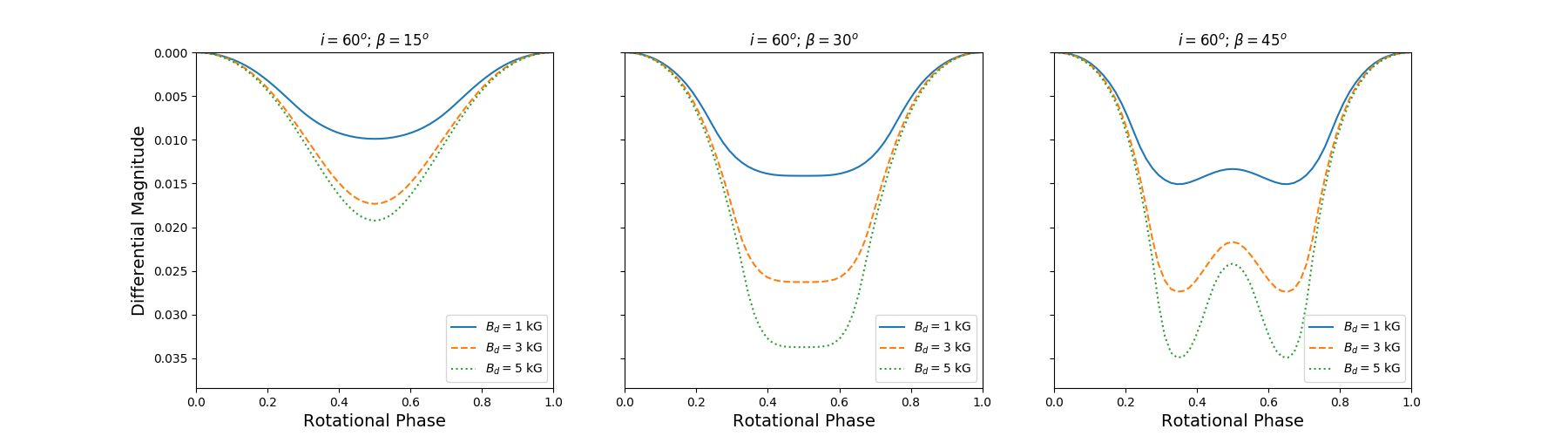}
	\caption{Model light curves with constant inclination. 
		The different panels show curves of increasing obliquity: $15^\circ$ (left), $30^\circ$(middle) and $45^\circ$ (right). Over-plotted on each panel are curves of increasing dipole field strength: 1 kG (solid), 3 kG (dashed) and 5 kG(dotted).  }
	\label{fig:params}
\end{figure}
From left to right, the panels in Fig. \ref{fig:params} illustrate the effect of increasing $i+\beta$ (for constant $B_d$). It can be seen that curves with $i+\beta=90^\circ$ are flat bottomed and mark the transition from single peaked curves ($i+\beta<90^\circ$) to double peaked curves ($i+\beta>90^\circ$). This is because at high $i+\beta$, the magnetic equator, corresponding to the densest region of the magnetosphere, crosses the observer's line-of-sight twice during one rotational cycle. Furthermore, plotted on each panel of Fig. \ref{fig:params}, we notice that curves of increasing $B_d$ with (constant $i+\beta$) will increase the depth of the light curve. This is to be expected as the Alfv\'{e}n radius will also increase,  consequently leading to greater occultation of the star's surface. However, at low $i+\beta$, the change in light curve depth is rather insensitive to $B_d$.

\section{Applications to observations}\label{sec:obs}

\subsection{HD 191612}
HD 191612 is a well-known Galactic Of?p star with observationally constrained magnetic field strength and field geometry. From longitudinal magnetic field measurements, a dipolar field strength of $B_d=2450\pm400$G was obtained satisfying the general set of solutions characterized by  $i+\beta=95\pm10^\circ$. While tentative Monte Carlo radiative transfer light curve models have been computed for this star, no genuine fit has ever been performed to its light curve. Using the same stellar parameters from \citet{Wade2011}, we attempt to fit a light curve to the Hipparcos photometry. The fitting procedure is accomplished via a Python implementation of the Markov chain Monte Carlo (MCMC) method using the \texttt{emcee} package from \citet{emcee}. We obtained $B_d=3.8_{-1.8}^{+0.9}$kG and $i+\beta=92_{-5}^{+5}$$^\circ$. It is reassuring to see that, within uncertainty, the best-fitted parameters are compatible with the previous findings obtained from direct measurements of the magnetic fields.

\subsection{LMC and SMC Of?p stars}
\begin{figure}
	\includegraphics[width=0.8\linewidth,trim={2cm 2cm 2cm 2cm}]{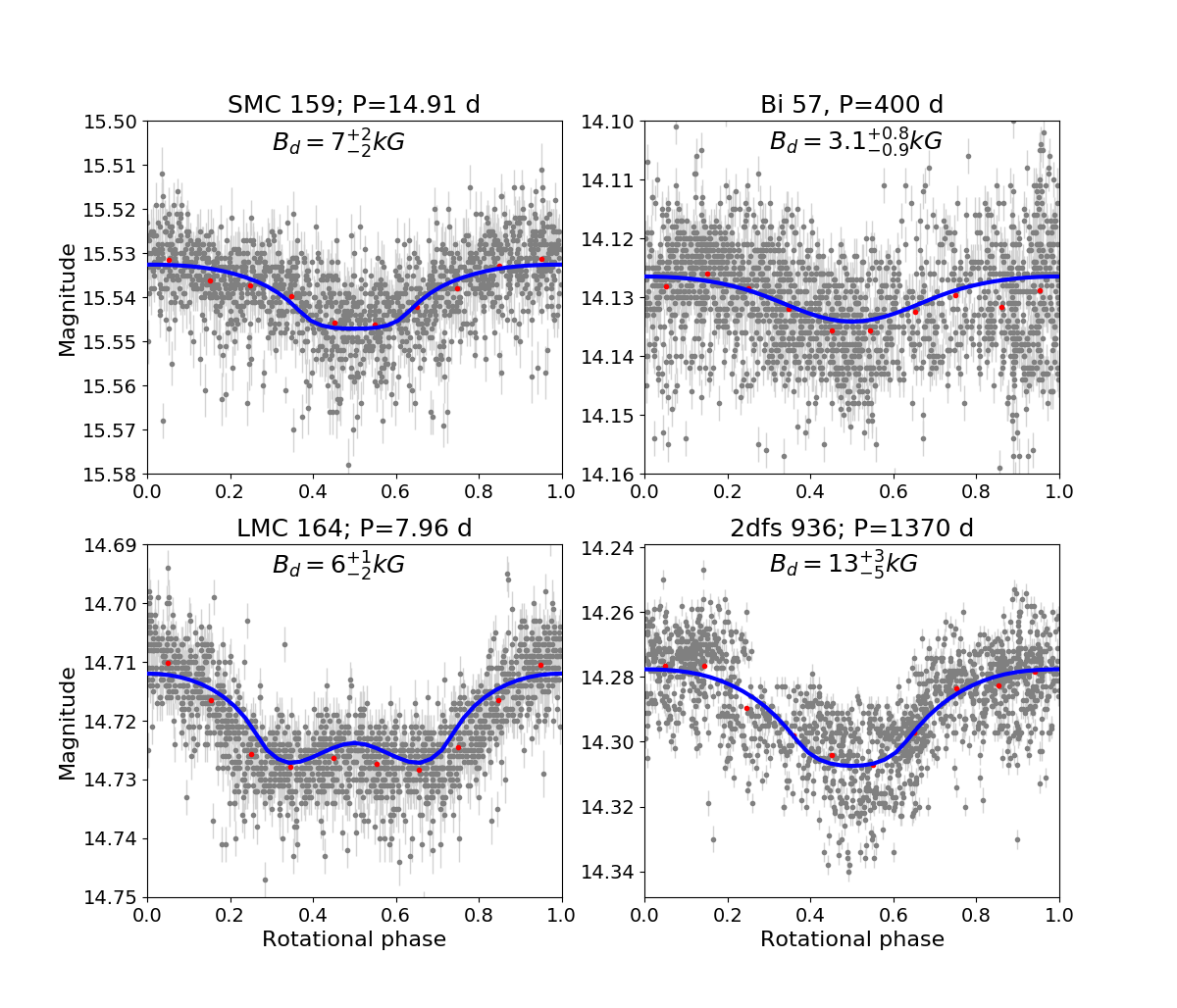}
	\caption{Phased OGLE light curves for the Of?p stars from the Magellanic Clouds. Bins corresponding to 0.1  cycles in phase are shown and best-fitted curves are over-plotted. }
	\label{fig:ogle}
\end{figure}

Among the five recently reported Of?p stars from the Magellanic Clouds \citep{Walborn,Naze}, only four stars have accurately determined photometric periods. Their phase-folded light curves are shown in Fig. \ref{fig:ogle}. We first utilized published prescriptions in order to deduce their wind parameters \citep{Vink2001}. The remaining stellar parameters were determined via spectroscopic modeling using FASTWIND \citep{FASTWIND}. The best-fit dipole field strengths from the MCMC fitting procedure are indicated on the top panels of Fig. \ref{fig:ogle}. We notice that the dipole field strengths are unusually high in comparison to the Galactic sample of Of?p stars. This is because the light curve depths from Magellanic Of?p stars are comparable to the Galactic Of?p stars, while the former have inherently lower mass-loss rates due their the sub-solar metallicities. It is therefore required for the dipole field strength to be increased in order to compensate for the reduced density in the magnetosphere.


\section{Discussion and conclusion}
To summarize, we described a simple prescription allowing for the light curve synthesis of magnetic massive stars. We apply this model to the OGLE photometry of the first candidate Of?p stars from the Magellanic clouds and predict the magnitude and geometry of their suspected magnetic fields. We note that these stars have already previously been observed with FORS2 \citep{Bagnulo}. However, due to unfavorable weather conditions and non-optimal sampling of observing times, no surface fields could be confirmed with certainty. We suspect that SMC 159, in particular, will be detectable with the FORS2 instrument and is scheduled to be re-observed in the coming year. 



\bibliographystyle{ptapap}
\bibliography{ptapapdoc}

\begin{thebibliography}{11}
\providecommand{\natexlab}[1]{#1}
\providecommand{\url}[1]{\texttt{#1}}
\providecommand{\urlprefix}{URL }
\providecommand{\eprint}[2][]{\url{#2}}

\bibitem[{{Bagnulo} et~al.(2017)}]{Bagnulo}
{Bagnulo}, S., et~al., \emph{{First constraints on the magnetic field strength
  in extra-Galactic stars: FORS2 observations of Of?p stars in the Magellanic
  Clouds}}, \emph{\aap} \textbf{601}, A136 (2017), \eprint{1703.00735}

\bibitem[{{Foreman-Mackey} et~al.(2013)}]{emcee}
{Foreman-Mackey}, D., et~al., \emph{{emcee: The MCMC Hammer}}, Astrophysics
  Source Code Library (2013), \eprint{1303.002}

\bibitem[{{Grunhut} et~al.(2017)}]{Grunhut}
{Grunhut}, J.~H., et~al., \emph{{The MiMeS survey of Magnetism in Massive
  Stars: magnetic analysis of the O-type stars}}, \emph{\mnras} \textbf{465},
  2432 (2017), \eprint{1610.07895}

\bibitem[{{Naz{\'e}} et~al.(2008){Naz{\'e}}, {Walborn}, \&
  {Martins}}]{Naze2008}
{Naz{\'e}}, Y., {Walborn}, N.~R., {Martins}, F., \emph{{The mysterious Of?p
  class and the magnetic O-star}}, \emph{\rmxaa} \textbf{44}, 331 (2008),
  \eprint{0807.3496}

\bibitem[{{Naz{\'e}} et~al.(2015)}]{Naze}
{Naz{\'e}}, Y., et~al., \emph{{Photometric identification of the periods of the
  first candidate extragalactic magnetic massive stars}}, \emph{\aap}
  \textbf{577}, A107 (2015), \eprint{1503.07654}

\bibitem[{{Owocki} et~al.(2016)}]{Owocki}
{Owocki}, S.~P., et~al., \emph{{An `analytic dynamical magnetosphere' formalism
  for X-ray and optical emission from slowly rotating magnetic massive stars}},
  \emph{\mnras} \textbf{462}, 3830 (2016), \eprint{1607.08568}

\bibitem[{{Santolaya-Rey} et~al.(1997){Santolaya-Rey}, {Puls}, \&
  {Herrero}}]{FASTWIND}
{Santolaya-Rey}, A.~E., {Puls}, J., {Herrero}, A., \emph{{Atmospheric
  NLTE-models for the spectroscopic analysis of luminous blue stars with
  winds.}}, \emph{\aap} \textbf{323}, 488 (1997)

\bibitem[{{Stibbs}(1950)}]{ORM}
{Stibbs}, D.~W.~N., \emph{{A study of the spectrum and magnetic variable star
  HD 125248}}, \emph{\mnras} \textbf{110}, 395 (1950)

\bibitem[{{Vink} et~al.(2001){Vink}, {de Koter}, \& {Lamers}}]{Vink2001}
{Vink}, J.~S., {de Koter}, A., {Lamers}, H.~J.~G.~L.~M., \emph{{Mass-loss
  predictions for O and B stars as a function of metallicity}}, \emph{\aap}
  \textbf{369}, 574 (2001), \eprint{astro-ph/0101509}

\bibitem[{{Wade} et~al.(2011)}]{Wade2011}
{Wade}, G.~A., et~al., \emph{{Confirmation of the magnetic oblique rotator
  model for the Of?p star HD 191612}}, \emph{\mnras} \textbf{416}, 3160 (2011),
  \eprint{1106.3008}

\bibitem[{{Walborn} et~al.(2015)}]{Walborn}
{Walborn}, N.~R., et~al., \emph{{Spectral Variations of Of?p Oblique Magnetic
  Rotator Candidates in the Magellanic Clouds}}, \emph{\aj} \textbf{150}, 99
  (2015), \eprint{1507.02434}

\end{thebibliography}

\end{document}